\documentclass[12pt]{article} 
\def\Bbb#1{{\bf #1}} 
\def\mathbb#1{{\Bbb #1}}
\makeatletter
 \@addtoreset{equation}{section}
 \makeatother

\begin{document}
\title{Octonionic M-theory and D=11 Generalized Conformal and 
Superconformal Algebras}

\author{Jerzy Lukierski${}^a$\thanks{Supported by KBN grant 
5P03B05620}
\thanks{{\em e-mail: lukier@ift.uni.wroc.pl}} 
and Francesco Toppan${}^b$\thanks{{\em e-mail: toppan@cbpf.br}}
\\ \\
${}^a${\it Institute for Theoretical Physics, University of
Wroc{\l}aw,}
\\ {\it 50-204 Wroc{\l}aw, pl. Maxa Borna 9, Poland}\\
 \\ ${}^b${\it CBPF, CCP, Rua Dr.}
{\it Xavier Sigaud 150,}
 \\ {\it cep 22290-180 Rio de Janeiro (RJ), Brazil}
}
\date{}
\maketitle
\begin{abstract}
Following [1] we further apply the octonionic structure to
supersymmetric D=11 $M$-theory.
 We consider the octonionic $2^{n+1} \times
2^{n+1}$ Dirac matrices describing the sequence of Clifford
algebras with signatures ($9+n,n$) ($n=0,1,2, \ldots$) and derive
the identities following from the octonionic multiplication table.
The case $n=1$ ($4\times 4$ octonion-valued matrices)
 is used for
the description of the $D=11$ octonionic $M$-superalgebra with
$52$ real bosonic charges; the $n=2$ case ($8 \times 8$
octonion-valued matrices)
 for the D=11 conformal
$M$-algebra with $232$ real bosonic charges. The  octonionic
structure is described explicitly for $n=1$ by the relations
between the $528$ Abelian $O(10,1)$ tensorial charges $Z_\mu$,
$Z_{\mu\nu}$, $Z_{\mu \ldots \mu_5}$ of the $M$-superalgebra. For
$n=2$ we obtain
 2080 real non-Abelian bosonic tensorial charges
 $Z_{\mu\nu}, Z_{\mu_1 \mu_2 \mu_3},
Z_{\mu_1 \ldots \mu_6}$ which, suitably constrained describe the
generalized $D=11$ octonionic conformal algebra. Further, we
consider the supersymmetric extension of this octonionic conformal
algebra which can be described as
$D=11$ octonionic superconformal
algebra
with a total number of $64$ real fermionic and $239$ real bosonic
generators.
\end{abstract}
\vfill{CBPF-NF-001/03}

\section{Introduction}

One of the consequences of the presence of membrane [2] and
five-brane [3] solutions of the $D=11$ supergravity is the 
appearance
in the $D=11$ superalgebra of $55$ two-tensor and $462$ five-
tensor Abelian
tensor charges, leading to the so--called $M$-
superalgebra\footnote{The
superalgebra (\ref{luto1.1}) is usually called $M$-algebra. We shall
use the terminology ``$M$-superalgebra" in order to point out that
it describes a supersymmetric theory.} [4--6] ($A,B = 1, \ldots ,
32; ~\mu,\nu = 0,1, \ldots, 10$)
\begin{equation}\label{luto1.1}
  \{Q_A, Q_B\} = Z_{AB} = (C\Gamma^\mu  )_{AB} P_\mu +
  (C\Gamma^{\mu_1\mu_2} )_{AB} \, Z_{[\mu_1\mu_2]}
  +(C\Gamma^{\mu_1\ldots \mu_5} )_{AB} \, Z_{[\mu_1\ldots \mu_5
]} \, .
\end{equation}
The $528$ real bosonic Abelian charges $Z_{AB}$ ($Z_{AB}  =
Z_{BA}$) saturate the rhs of the anticommutator of $32$ real
supercharges $Q_A$, which are $D=(10,1)$ as well as $D=(10,2)$
Majorana spinors. In $D=12$ the $M$-superalgebra (\ref{luto1.1})
takes the form [7,8] ($A,B=1, \ldots 32;~ \widetilde{\mu},
\widetilde{\nu} \ldots = 0,1, \ldots,11$)
\begin{equation}\label{luto1.2}
  \{ Q_A, Q_B \} = Z_{AB} =
  \left(C\Gamma^{\widetilde{\mu}_1\widetilde{\mu}_2}\right)_{AB}\,
  Z_{[\widetilde{\mu}_1\widetilde{\mu}_2]}
  +
  \left(C\Gamma^{\widetilde{\mu}_1 \ldots 
\widetilde{\mu}_6}\right)_{AB}\,
  Z_{[\widetilde{\mu}_1 \ldots \widetilde{\mu}_6]}
  \, ,
\end{equation}
where $Z_{[{\mu}11]}= P_\mu$ and the five-tensor charges in
(\ref{luto1.1}) are described by
 D=12 six--charges
 $Z_{[\widetilde{\mu}_1 \ldots
\widetilde{\mu}_6]}$
 satisfying
  the self-duality condition.

Let us recall that $O(10,2)$ describes the standard $D=10$
conformal algebra, with fundamental spinor ($D=10$ twistor) built
from a pair of $16$-components $D= (9,1)$ Majorana--Weyl spinors.
It appears that these $16$ components can be endowed with
octonionic structure, and we can describe $D=10$ spinors as a pair
of octonions $\pmatrix{U_1\cr U_2}$. Subsequently one can
introduce the following relation of the octonionic matrices with
D=10 Lorentz and conformal groups [9--12]
\begin{equation}\label{luto1.3}
  SL(2|{\bf{O}}) = \frac{SO(9,1)}{G_2}\, , \qquad
  U_\alpha (4|{\bf{O}}) = \frac{SO(10,2)}{G_2} \, ,
\end{equation}
where the $14$-generator $G_2$ algebra is the automorphism 
algebra of
the multiplication table for the octonionic units $t_k$ ($k=1
\ldots 7$)
\begin{equation}\label{luto1.4}
  t_k t_l = - \delta_{kl} +\frac{1}{2} \, f_{kl}^m \,t_m
\end{equation}
and $U_\alpha (n|{\bf{F}})$ describes the antiunitary
${\bf{F}}$-valued group, with unitary norm and antisymmetric
 invariant metric, where
${\bf{F}}={\bf{R}},{\bf{C}},{\bf{H}},{\bf{O}}$. For $n=4$ we
obtain the sequence of $D=3$ (${\bf{F}}={\bf{R}}$), $D=4$
(${\bf{F}}={\bf{C}}$) and $D=6$ (${\bf{F}}={\bf{H}}$) spinorial
coverings of conformal groups; for ${\bf{F}}={\bf{O}}$ we obtain
the second formula (\ref{luto1.3}). Subsequently we shall denote
by $U_\alpha(n|{\bf F})$ also the corresponding algebras, with
${\bf F}$--Hermitean generators.

Using the first relation in (\ref{luto1.3}) it has been
proposed
[13] that the standard $D=10$ Poincar\'{e} superalgebra
can be described by a pair of octonionic supercharges
(${\cal{Q}}_1, {\cal{Q}}_2$), with the following basic relations
(${\cal{Q}} = Q^{(0)} + t_r Q^{(r)} \to
{\cal{Q}}^\dagger\equiv{\overline{\cal{Q}}} = Q^{(0)} -t_r Q^{(r)}
$)
\begin{equation}\label{luto1.5}
  \{ {\cal{Q}}_{a}, {\overline{\cal{Q}}}_b \} = {\cal{Z}}_{ab} =
  \pmatrix{P_0 +P_9 & P_8+t_rP_r\cr
  P_8-t_rP_r & P_0-P_9 }\, ,
\end{equation}
where
 $Z_{ab} = Z^{+}_{ba}$ and
  $P_\mu = (P_0,P_1,\ldots P_9)$ describe $D=10$ momentum
generators.

Recently we proposed also to impose
 on the $M$-superalgebra generators
 (see  formulae (\ref{luto1.1}) and (\ref{luto1.2})) the
octonionic structure [1]. The relations
(\ref{luto1.1}--\ref{luto1.2}) we replaced by octonionic
$M$-superalgebra ($r,s=1,2,3,4$)
\begin{equation}\label{luto1.6}
  \{ {\cal{Q}}_r, {\overline{\cal{Q}}}_s \} = {\cal{Z}}_{rs} \, ,
  \qquad \qquad {\cal{Z}}_{rs} = {\cal{Z}}^+_{sr}\, ,
\end{equation}
where
four real octonion-valued supercharges $Q_r$ replace $32$
 real supercharges $Q_A$ and
  in place of the $528$ real Abelian charges present in the rhs of
(\ref{luto1.1}--\ref{luto1.2}) we get only $ 52$ independent real
Abelian supercharges described by the real components 
($Z^{(0)}_{rs},
Z^{(k)}_{rs}$) where
\begin{eqnarray}\label{luto1.7}
{\cal{Z}}_{rs}& = & Z^{(0)}_{rs}+ t_k Z^{(k)}_{rs}\, ,
\\
Z^{(0)}_{rs} &= & Z^{(0)}_{sr}
\, , \qquad
 Z^{(k)}_{rs} = - Z^{(k)}_{sr} \, .
\end{eqnarray}
As a consequence, the octonionic $M$-algebra can be fully
described by $11$ four-momenta $P_\mu$ and
 only  $41$ generators
${\cal Z}_{\mu\nu}$ describing together the Abelian contractions
of all the generators in the  coset $\frac{O(10,2)}{G_2}$; another
way of providing the bosonic generators of (\ref{luto1.6}) is to
introduce suitably constrained five-tensor charges ${\cal
Z}_{[\mu_1\ldots \mu_5]}$.

In this paper we continue the considerations presented in [1].
In Sect. {\bf 2} we shall consider more in detail the properties of the
octonionic $M$-superalgebra (\ref{luto1.6}) and present it as a
member of generalized octonionic supersymmetry algebras in $
D=(9+n,n)$ ($n=0,1,2, \ldots $). For $n=0$ we obtain the
 generalized supersymmetry algebra for
  $D=9$ Euclidean
theory. If $n=1$ we get  the octonionic $M$-superalgebra
  considered in [1],
 and the case $n=2$ provides
       the extension of $M$-superalgebra
 to $D=13$ with signature
$(11,2)$ and $D=14$ with signature $(11,3)$, considered by
Bars [14,15]\footnote{Such a framework was proposed
 as a basis of the so--called $S$-theory
unifying the algebraic description
of all five superstring theories in $D=10$.
 Similarly as the
$M$-theory which can be described in the $(10,2)$--covariant
framework, the $S$--theory can be as well described by the $D=14$
formalism, with $(11,3)$ signature [13].}. It appears however that
the relation (\ref{luto1.6}) if $r,s,=1,2,\ldots, 2^n$, can also be
used for the supersymmetrization of the octonionic algebras
$U_\alpha(2^n|{\bf O})$, which describe for $n=1$ the octonionic
$D=(9,1)$ algebra with ten curved translations, for $n=2$ the
octonionic conformal algebra (see ($1.3$)) and for $n=3$ the
$D=11$ octonionic generalized conformal
 algebra $U_{\alpha}(8|{\bf O})$
  with $232$ real
charges. We shall argue that $U_\alpha(8|{\bf O})$ can be
 obtained from
  the generalized $D=11$ conformal algebra
$Sp(64|{\bf{R}})$
 [16,17]  by imposing
 constraints describing the  octonionic structure. It
is known [17]
 that the $Sp(64|{\bf R})$ generators can be described by $O(11,2)$
two--tensors, three--tensors and six--tensors.
We shall show that the octonionic conformal $M$--algebra
 $U_{\alpha}(8|{\bf O})$
 will be
completely described only in terms of the two--tensor generators
from the coset $\frac{O(11,2)}{G_2}$
 supplemented by
 a suitably
restricted set of three--tensor generators.
 It appears that $U_\alpha(8|{\bf O})$ supplemented with $S^7 
\simeq U(1|{\bf O})$
 generators describe the bosonic sector of
  the supersymmetric extension $U_\alpha(8;1|\bf {O})$
of the octonionic conformal $M$--algebra.

 The conformal algebras in
$D=10$, since fundamental
D=10
  octonionic conformal spinors  have  four components,
 belong to  the framework of conformal
   Jordan
algebras
 [9--11,18,12]). Beyond $D=10$ and in particular for the eleven
dimensional conformal  $M$-algebra $U_\alpha (8;{\bf O})$, we are 
outside
 of the framework of conformal algebras
associated with
  Jordan algebras. The construction of the conformal algebra
 can be however linked with
 the group of invariance of the metric for
``doubled Lorentz spinors".  This procedure we propose to apply to
octonionic conformal spinors in dimensions $D=(9+n,n)$, providing 
$D=11,13$
etc. We consider in Sect. 3 the generalized octonionic conformal
algebras and superalgebras in the form of
  octonionic-valued (super-)matrix
realizations  which admit closed algebraic relations with the
(super-)matricial (anti-)commutator structure.
For D=11, considering
  $8\times 8$ octonionic matrices $U_{\alpha}(8|{\bf O}) \simeq
 Sp(8|{\bf O})$, one can show that it contains the generators
 parametrizing the coset
  $\frac{O(11,2)}{G_2}$
  ($64$ real generators), but with additional $168$ real generators
   not closing to any subset of Lie algebra generators. Such
algebra is also not of Malcev type [20]; it is an interesting
task to elucidate its algebraic characterization.

Further, in Sect. 4,  we provide the table listing the number of
independent $n$-fold antisymmetric  products of octonionic Dirac
matrices in odd dimensions $D=7+2k$ ($k=0,1,2,3$) and we
interprete them as the relation between the $p$-brane   degrees of
freedom in corresponding supersymmetric
 $D$-dimensional theory with the most general set of central
 charges.

\section{Octonionic-Valued Clifford Algebras and Corresponding
Space--Time Superalgebras}

\hskip15pt {\bf  a) D=(0,7)}

Let us observe that (0,7) spinors are real eight--dimensional and
$C= C^T$\footnote{In all the examples below the $C$ matrix is 
given by the product of
the time-like Clifford's Gamma matrices.}. Introducing seven 
$8\times 8$
real $\Gamma_i$
matrices, $i=1, \ldots, 7$,
\begin{equation}\label{luto2.1}
  \{ \Gamma_i, \Gamma_j \} = -2 \delta_{ij} \, ,
\end{equation}
one obtains the relations
($\Gamma_{i_1 \ldots i_n}
 = \frac{1}{n}
  \sum\limits_{\rm perm}
   (-1)^{\rm perm}
  \Gamma_{i_{ k_1}} \Gamma_{i_{ k_1}}\ldots \Gamma_{i_{ k_n}}$)
\begin{equation}\label{luto2.2}
  C\Gamma_i = - (C\Gamma_i)^T \, , \quad
C\Gamma_{ij} = - (C\Gamma_{ij})^T \, , \quad C\Gamma_{ijk} = -
(C\Gamma_{ijk})^T  \, .
\end{equation}
If we use 8 real supercharges one obtains the D=7
 generalized
 Euclidean
superalgebra ($\alpha = 0,1,2,\ldots 7$)
\begin{equation}\label{luto2.3}
  \{Q_\alpha , Q_\beta \} = Z \cdot (C)_{\alpha \beta} + Z^{ijk}
  (\Gamma_{ijk})_{\alpha\beta} \, ,
\end{equation}
which unfortunately does not contain  
the
7-momentum sector which should be linear in $\Gamma_i$.

One obtains the octonionic $C_{\bf O}(0,7)$ Clifford algebra with
generators ${\Gamma_i}^{(7)}$ satisfying the relation
(\ref{luto2.1}) by assuming
\begin{equation}\label{luto2.4}
  \Gamma_{i}^{(7)} = t_i \, ,\qquad C = 1\, ,
\end{equation}
where $t_i$ are seven octonionic units with the
 multiplication table (\ref{luto1.4}). The Hermitean $N=1$ octonionic
superalgebra generated by the supercharge ${\cal{Q}} = Q_0 + t_a 
Q_a$
\begin{equation}\label{luto2.6}
  \{ \overline{\cal{Q}}, {\cal{Q}} \} = {\cal {Z}} \, ,
\end{equation}
describes the octonionic $N=8$ supersymmetric mechanics, with 
the real
generator ${\cal Z}$ playing the role of the Hamiltonian [18].

\vskip12pt

 {\bf b) D= (1,8),(1,9) and D=(9,0),(9,1).}

\vskip8pt

One can introduce the following $2\times 2$ matrix realizations
of the octonionic--valued Clifford algebras:

\vskip12pt

{\sl b1) The $C_{ {\bf{O}}} (1,8)$ Clifford algebra ($C^T =
-C$)}

\vskip8pt

\begin{equation}\label{luto2.7}
  \Gamma_i^{(9)} = 
\left( \begin{tabular}{c|c}
0 & $t_i$ \cr \hline
$t_i$ & 0 \end{tabular}\right)
  \, , \quad   \Gamma_8^{(9)}=\pmatrix{1 & 0 \cr 0 &-1 } \, ,
\quad \Gamma_0^{(9)} = \pmatrix {0 &1\cr -1& 0}\, ,
\end{equation}
where $C= \Gamma_0$. The generalized octonionic D=9 
Poincar\'{e}
algebra takes the form ($r,s = 1,2; \mu = 0,1 \ldots 8$)
\begin{equation}\label{luto2.8}
  \{ {\cal{Q}}_r, \overline{{\cal{Q}}}_s\} = C_{rs} {Z} + (C 
\Gamma_\mu) _{rs} \,P^\mu \, .
\end{equation}
We see that the generator $Z$ in D=9 is the central charge. Due
however to the first relation (\ref{luto1.3}) the superalgebra
(\ref{luto2.8}) can be promoted to D=(1,9) superPoincar\'{e}
algebra [10], where now the $D=9$ central charge $Z$ is the
component $P^9$ of the ten--dimensional momenta. One can also
assume that the generators on the rhs of ($2.7$) are nonAbelian
and form the algebra $U_\alpha(2|{\bf O})$ containing only ten
curved ($1.9$) translations, which is an octonionic counterpart of
the $D=10$ AdS algebra. \vskip12pt

 {\sl b2) The $C_{\bf{O} } (9,0)$ Clifford algebra ($C^T =
C$)}

\begin{equation}\label{luno2.9}
  \widetilde{\Gamma}^{(9)}_i = \pmatrix{0 & t_i\cr -t_i & 0}\,
  ,\qquad
   \widetilde{\Gamma}^{(0)}_i = \pmatrix{0 & 1\cr 1 & 0}\,
  ,\qquad
   \widetilde{\Gamma}^{(8)}_i = \pmatrix{1 & 0\cr 0 & -1}\,
  ,
\end{equation}
where $C= {{1\!\!1}}_2$. We obtain
  the
 general octonionic
   $D=9$ Euclidean superalgebra
with $D=9$ Euclidean momenta $\widetilde{P}_\mu$   ($\mu=1 
\ldots
9$) and central charge $ \widetilde{Z}$ ($r,s=1,2$)
\begin{equation}\label{luto2.10}
  \{ {\cal{Q}}_r , \overline{{\cal{Q}}}_s \} = \delta_{rs} \widetilde{Z} +
  (\widetilde{\Gamma_\mu})_{rs} \widetilde{P}^\mu \,  .
\end{equation}
Again, the superalgebra (\ref{luto2.10}) can be promoted to 
$D=(9,1)$
Poincar\'{e} algebra if we choose $ \widetilde{Z}= P_0$ (the $D=10$
energy operator).

\vskip12pt

{\bf  c) ${\bf D= (10,1),(10,2)}$}

\vskip12pt

{\sl c1) The $C_{\bf{O}} (10,1)$ Clifford algebra and D=11
 $M$-superalgebra.}

\vskip8pt
Such an algebra can be
 represented by the following $4\times 4$ octonionic matrices%
\footnote{One can represent
 $C_{\bf O} (10,1)$ in two ways.
  The second choice is obtained when the
  following $D=11$ octonionic $4\times 4$ Dirac matrices are taken
(now $C=\widetilde{\Gamma}^{(11)}_{0}$)
\begin{equation}\label{luto2.15}
  \widetilde{\Gamma}^{(11)}_{a} =
  \pmatrix{0 &\widetilde{\Gamma}^{(9)}_{b}\cr
  \widetilde{\Gamma}^{(9)}_{b} & 0} \, ,
   \quad
 \widetilde{\Gamma}^{(11)}_{0} =
  \pmatrix{0 & {\bf 1}_2\cr
 -{\bf 1}_2& 0} \, , \quad
\widetilde{\Gamma}^{(11)}_{10} =
  \pmatrix{{\bf 1}_2 & 0 \cr
  0 & -{\bf 1}_2} \, .
\end{equation}.}
  ($a=b+1= 1, \ldots, 9$)

\begin{equation}\label{luto2.11}
  \Gamma^{(11)}_{a} = \pmatrix{0 & \Gamma^{(9)}_b \cr -
  \Gamma^{(9)}_b &  0} \, , \quad
   \Gamma^{(11)}_{0} = \pmatrix{0 & {\bf 1}_2 \cr
  {\bf 1}_2 &0}\, , \quad
   \Gamma^{(11)}_{10} = \pmatrix{{\bf 1}_2 & 0 \cr
  0 & - {\bf 1}_2}\, ,
\end{equation}
where \begin{equation}
C^{(11)}= \pmatrix{0 & \Gamma^{(9)}_8\cr -
  \Gamma^{(9)}_8 &  0}.\label{luto2.11bis}
  \end{equation}

Introducing $4$ octonionic supercharges we obtain the octonionic
$M$-su\-per\-al\-geb\-ra (\ref{luto1.6}) considered in our
previous paper [1].
The superalgebra (\ref{luto1.6}) has $52$ independent real bosonic
charges, which can be equivalently expressed in two ways
 ($r,s=1, \ldots 4$)
\begin{equation}\label{luto2.12}
  {\cal{Z}}_{rs} = P^\mu (C\Gamma^{(11)}_\mu)_{rs} +
   Z^{\mu\nu}_{\bf{O}} (C\Gamma^{(11)}_{\mu\nu})_{rs} =
    Z^{[\mu_1\ldots \mu_5]}_{\bf{O}}
    (C\Gamma^{(11)}_{\mu_1 \ldots
    \mu_5})_{rs}\, ,
\end{equation}
where all $11$ components of $P^k$ are independent, while the 
generators
$Z^{\mu\nu}_{\bf{O}}$ describe the coset $\frac{O(10,1)}{G_2}$
(i.e. there are $14$ relations between the $55$ generators
$Z^{\mu\nu}_{\bf{O}}$), and out of the $462$ components of 
$Z^{[\mu_1
\ldots \mu_5]}_{\bf{O}}$ there are only $52$ independent ones.

Similar constraints are satisfied by the coordinates of the
generalized octonionic $D=11$ space-time
\begin{equation}\label{luto2.14}
  {\cal{X}}_{rs} = X^{\mu} (C\Gamma^{(11)}_{\mu} )_{rs} +
  X^{\mu\nu}_{\bf{O}} (C\Gamma^{(11)}_{\mu\nu} )_{rs}=
  X^{\mu_1 \ldots \mu_5}_{\bf O} (C\Gamma^{(11)}_{\mu_1 \ldots 
\mu_5}
  )_{rs}\, .
\end{equation}
We see that in the $D=11$ supersymmetric theory with basic
superalgebra (\ref{luto2.12}) the two-brane and five-brane
tensorial central charges are strongly constrained, and the
 theory can be described
 completely

 - either by one-brane and two-brane sectors
 $((P^{\mu}, Z^{\mu\nu}_{\bf{O}})$ in generalized momentum
  space or, using the dual picture, by
  $(X^{\mu}, X^{\mu\nu}_{\bf{O}})$ in generalized space-time,

  - or by the constrained five-brane sector $(Z^{\mu_1 \ldots 
\mu_5}_{\bf O}$
  or $X^{\mu_1 \ldots \mu_5}_{\bf{O}})$.

\vskip12pt

{\it c2) The $D=10$ octonionic conformal superalgebra $U_{\alpha}
(4|{\bf O})$}

\vskip8pt

The form  (\ref{luto1.6}) of the octonionic superalgebra is also
obtained for the generalized octonionic $D=10$ conformal
superalgebras. For that purpose one can  write the basic
relation in $O(10,2)$ - covariant form
 ($r,s=1,\ldots, 4; \widetilde{\mu}_1,\widetilde{\mu}_2 = 0,1,\ldots 
11$)
\begin{equation}\label{luto2.16}
  \{ \widetilde{{\cal{Q}}}_r, \overline{\widetilde{{\cal{Q}}}}_s \} = 
{\cal{M}}_{rs} =
  (C\Gamma^{(12)}_{ \widetilde{\mu}_1 \widetilde{\mu}_2})_{rs} \,
  M^{[\widetilde{\mu}_1 \widetilde{\mu}_2]}\, ,
\end{equation}
where (${\mu}_1, {\mu}_2 = 0,1, \ldots 10)$

\begin{equation}\label{luto2.17}
\Gamma^{(12)}_{{\mu}_1 {\mu}_2}
= \Gamma^{(11)}_{{\mu}_1 {\mu}_2}
\equiv \left[ \Gamma^{(11)}_{
{\mu}_1 }\,, \Gamma^{(12)}_{  {\mu}_2}\right]\,, \qquad
\Gamma^{(12)}_{ 11 {\mu} } = \Gamma^{(11)}_{ {\mu}}\, ,
\end{equation}
with the $66$ generators 
$M_{[\widetilde{\mu}_1,\widetilde{\mu}_2]}$
describing $52$ independent generators of $\frac{O(10,2)}{G_2}$ 
(see
(1.3b)). In the octonionic framework the remaining
 six-tensor generators $M_{[\widetilde{\mu}_1 \ldots
\widetilde{\mu}_6]}$
 (compare with (1.2)) are not
  independent.
The octonionic superalgebra (2.15) can be also treated as
describing D=11 octonionic $AdS$ algebra [10],
 with $\relax M^{[\mu_1\mu_2]}$ describing $\frac{O(10,2)}{G_2}$
and $M^{12\mu}$ the curved $D=11$ AdS translations.
 In particular if we
introduce the rescaling of the generators
 $M^{12\mu} = R \cdot {\cal P}^{\mu}$
 by performing
the limit
 $R\to \infty$ one can obtain from  the relations (2.15) the
  D=11 octonionic $M$-algebra, given by (1.6).
\par We add here that by doubling the realization of $C_{\bf
O}(10,1)$ given in footnote $~^{4)}$ one obtains the realizations
of the Clifford algebras $C_{\bf O}(2,9)$ and $C_{\bf O}(2,10)$.

\vskip12pt

{\bf d)}  {\bf D=(11,2) }

In D=13 we shall consider only one choice of signature
(11,2)\footnote{The other possible choice of octonionic $8\times
8$ Clifford algebra representations can be considered for
signatures $(3,10)$ and $(3,11)$.}. The corresponding
representation of $C_{\bf O}(11,2)$ in terms
   of $8\times 8$ octonionic matrices is
  explicitly given, for
($a=b+1= 1, \ldots, 11$), as follows:
\begin{equation}\label{luto3.1}
  \Gamma^{(13)}_{a} = \pmatrix{0 & \Gamma^{(11)}_b \cr
  \Gamma^{(11)}_b &  0} \, , \quad
   \Gamma^{(13)}_{0} = \pmatrix{0 & {\bf 1}_2 \cr
  -{\bf 1}_2 &0}\, , \quad
   \Gamma^{(13)}_{12} = \pmatrix{{\bf 1}_2 & 0 \cr
  0 & - {\bf 1}_2}\, ,
\end{equation}
with $C^{(13)}= \pmatrix{C^{(11)} & 0\cr
  0 &  -C^{(11)}}$, while $\Gamma^{(11)}_b$ and $C^{(11)}$ are
  given by (\ref{luto2.11}) and (\ref{luto2.11bis}) respectively.

The basis of the $232$ octonionic
  hermitian generators is given by the $64$
antisymmetric two-tensors
$C\Gamma_{[\mu_1\mu_2]}$ 
and the $168$
antisymmetric three tensors $ C\Gamma_{[\mu_1\mu_2\mu_3]}$
or, equivalently, by the $232$
antisymmetric six-tensors $ C\Gamma_{[\mu_1\ldots \mu_6]}$.

The D=(11,2) generalized octonionic supersymmetry algebra
(called also $S$-algebra [14,15]) takes the form
 ($r,s=1,2 \ldots 8; M,N=0,1,2,\ldots 12$)
 \begin{equation}
\left\{ Q_r , Q_s \right\} =
\left(
C^{(13)} \, \Gamma^{(13)}_{MN}
\right)_{rs} \, Z^{MN}_{\bf 0} +
\left(
C^{(13)}    \, \Gamma_{M_1 M_2 M_3}
\right)_{rs}
Z^{M_1 M_2 M_3}_{\bf 0} \, , 
 \end{equation}
with $232$ real bosonic charges, which is an octonionic
counterpart of the real generalized D=(11,2) superalgebra with
$64$ real Majorana  supercharges ($A,B = 1 \ldots 64$) and $2048$
bosonic charges.
 \begin{eqnarray}
\left\{ Q_A , Q_B \right\}
&= &
\left(
C^{(13)} \, \Gamma^{(13)}_{MN}
\right)_{AB} \, Z^{MN} +
\left(
C^{(13)}    \, \Gamma_{M_1 M_2 M_3}
\right)_{AB}
Z^{M_1 M_2 M_3}
\cr
&&
 - \
\left( C^{(13)} \, \Gamma_{M_1 \ldots M_6} \right)_{AB}
Z^{M_1 \ldots M_6}
\, ,
 \end{eqnarray}
where now $(\Gamma^{(13)}_M)_{AB}$ are the $64\times 64$ real
Majorana   representations of $C(11,2)$.
>From the formula (2.20)  follows that

i) The octonionic six-charges are not needed in the relation
 (2.18) because they
  can be expressed by two-charges and three-charges.

 ii) The relations (2.19) with non Abelian charges $Z^{MN},
 Z^{M_1 M_2 M_3}, Z^{M_1 \ldots M_6}$ describe the superalgebra
  $OSp(1|64;R)$, which was proposed as the generalized D=11 
conformal
  superalgebra [17,19].
In such a case the generators $Z^{MN}$ describe
  the $O(11,2)$  algebra. The octonionic counterpart of
   $Sp(64;R)$ is provided by $U _\alpha(8| {\bf O})$, which is
   supersymmetrized by relation (2.18). The generators
    $Z^{MN}_{\bf 0}$ describe in such a case the coset
    $\frac{O(11,2)}{G_2}$.

\section{D=11 Octonionic Generalized  (Super)conformal
Transformations as Automorphisms}

It is known that the conformal algebra can be introduced as the
algebra of transformations leaving invariant the inner product of
fundamental conformal
  spinors called also twistors.
  We shall apply this method to derivation of octonionic conformal 
algebra from
   octonionic spinors with inner product.
   In $D=(10,1)$ such inner  product  is given by
$\psi^\dagger C \eta$, where
 $\psi, \eta$ are eightdimensional octonionic
  conformal $O(11,2)$ spinors
  described by pairs of octonionic $O(10,1)$ Lorentz spinors
   and
the matrix $C$
  given by the product of the two space-like Clifford's
Gamma matrices $\Gamma^{(13)}_{ 0},
\Gamma^{(13)}_{12}$ (see (2.17))
 is real-valued and totally antisymmetric.
Therefore, the conformal transformations are realized by the
octonion-valued $8$-dimensional matrices ${\cal M}$ leaving $C$
invariant, i.e. satisfying
\begin{eqnarray}
{\cal M}^\dagger C + C {\cal M} &=& 0.
\end{eqnarray}
This allows identifying the
octonionic   conformal
transformations with the
octonionic unitary-symplectic
transformations $U_\alpha(8|{\bf O})$.

The most general octonionic-valued matrix leaving invariant
$\Omega$ can be expressed as follows
\begin{eqnarray}
{\bf M} &=&\left(
\begin{array}{cc}
D& B \\ C & -D^\dagger
\end{array} \right),\label{confM}
\end{eqnarray}
where the $4\times 4$ octonionic matrices $B$, $C$ are hermitian
\begin{eqnarray}
B=B^\dagger,\quad &&\quad C=C^\dagger .\label{BCcond}
\end{eqnarray}
It is easily seen that the total number of independent components
in (\ref{confM}) is precisely $232$, as we expected from the
previous considerations.

It should be noticed that the set of octonionic
  matrices ${\bf M}$ of
(\ref{confM}) type forms a closed algebraic structure under the
usual matrix commutation. Indeed one gets
$\relax [ {\bf M}, {\bf
M}] \subset {\bf M}$, endowing the structure of $U_\alpha(8|{\bf
O})$ to ${\bf M}$. As recalled in the introduction,
 $U_\alpha(2n;{\bf O})$
 for $n>3$  is no longer a
conformal algebra associated with a Jordan-algebra (see e.g
\cite{gpr}), nevertheless
  it admits
  the Lie-algebraic commutation relations which, however,
  do not satisfy the Jacobi identities.

In the procedure of supersymmetric extension to the
superconformal algebra we have to accommodate the components 
of
$8$ octonionic spinors of $(11,2)$ into a supermatrix enlarging
$U_\alpha(8|{\bf O})$. This can be achieved as follows. The two
$4$-column octonionic spinors $\alpha$ and $\beta$ can be
accommodated into a supermatrix of the form
\begin{eqnarray}&&
{\cal M}^{(1)} \ = \
\left(\begin{array}{c|cc}
  0 & -\beta^\dagger & \alpha^\dagger\\ \hline
  \alpha & 0& 0 \\
  \beta & 0 & 0
\end{array}\right)\label{fermionic}.
\end{eqnarray}
Under anticommutation, the lower bosonic diagonal block reduces 
to
$U_\alpha(8|{\bf O})$. On the other hand, extra real seven 
generators,
associated to the $1$-dimensional antihermitian matrix $A$
\begin{eqnarray}
A^\dagger &=& - A, \label{Acond}
\end{eqnarray}
i.e. described by  seven imaginary octonions, are obtained in
the upper bosonic diagonal block. Therefore, the generic bosonic
element is of the form
\begin{eqnarray}&&
{\cal M}^{(0)} \ = \
\left(\begin{array}{c|cc}
  A & 0 & 0\\ \hline
  0 & D& B \\
  0& C & -D^\dagger
\end{array}\right)\label{bosonic},
\end{eqnarray}
with $A$, $B$ and $C$ satisfying (\ref{Acond}) and
(\ref{BCcond}).\par

It can be shown that in analogy to the relation (3.1) one can
 derive (3.6) from the invariance of inner product for octonion-valued
  supertwistor ($\psi, \xi$), where $\xi = \xi^{(0)} +
  \xi^{(i)}t_i$
  ($\{ \xi^{(a)}, \xi^{(b)}\} = 0; a,b=0,1, \ldots 7$ and $\xi^{(a)}$
  real):
\begin{equation}\label{mmmcc}
{\cal M}^{\#} \mathbb{C} = - \mathbb{C} \, {\cal M}^{\#}, ,
 \qquad
 {\cal M}  = {\cal M}^{(0)} \oplus {\cal M}^{(1)}\, ,
\end{equation}
where $\mathbb{C} = \pmatrix{1 & 0\cr 0 &C}$ is the $OSp(1;8)$
metric and $\#$ describe graded-Hermitean conjugation. The closed
superalgebraic structure, with (\ref{fermionic}) as generic
fermionic element and (\ref{bosonic}) as generic bosonic element,
we denote as $OSp(1,8|{\bf O})$. It can be considered as the
superconformal algebra of the octonionic $M$-theory or generalized
octonionic AdS algebra in $D=(11,1)$ and admits a total number of
$239$ real bosonic generators.

\section{Octonionic Structure and $p$-Superbranes }

We have shown that the $52$ independent components of the
 Hermitian-octonionic
$Z_{ab}$ matrix can be represented either as the $11+41$ bosonic
generators entering
\begin{equation}\label{eq1}
  {\cal{Z}}_{ab} = P^\mu (C\Gamma^{}_\mu)_{ab} +
   Z^{\mu\nu}_{\bf{O}} (C\Gamma^{}_{\mu\nu})_{ab}
   ,
\end{equation}
or as the $52$ bosonic generators entering
\begin{equation}\label{eq2}
  {\cal{Z}}_{ab} =
    Z^{[\mu_1\ldots \mu_5]}_{\bf{O}}
    (C\Gamma^{}_{\mu_1 \ldots
    \mu_5})_{ab}\, .
\end{equation}
The reason for that lies in the fact that, unlike in the real case,
the sectors individuated by (\ref{eq1}) and (\ref{eq2}) are not
independent.
This is a consequence of the multiplication    table
 of the
 octonions. Indeed,  when we  multiply
antisymmetric products of $k$ octonionic-valued Gamma matrices, 
a
certain number of them are redundant. For $k=2$, due to the $G_2$
automorphisms, $14$ such products have to be erased. In the
general case \cite{crt} a table can be produced, which
 we write down (see Table $1$)
for $7\leq D \leq 13$
 odd-dimensional spacetime corresponding to octonionic 
realizations of
Clifford algebras considered in Sect. 2.
  The  Table~1 was  constructed from the
$D=7$ results (which can be easily computed), by taking into
account that out of the $D$ Gamma matrices, $7$ of them are
octonion-valued, while the remaining $D-7$ are purely real. The
columns in Table $1$ are labeled by $k$, the number of
antisymmetrized Gamma matrices. \vfill\eject
\begin{table}[h]
\begin{center}
\begin{tabular}{|c|c|c|c|c|c|c|c|c|c|c|c|c|c|c|}\hline
$ 
$&$0$&$1$&$2$&$3$&$4$&$5$&$6$&$7$&$8$&$9$&$10$&$11$&
$12$&$13$\\
\hline
$D=7$&$\underline{1}$&$7$&$7$&$\underline{1}$&$\underline{1}$&
$7$&$7$&$
\underline{1}$&&&&&&\\ \hline
$D=9$&$\underline{1}$&$\underline{9}$&$22$&$22$&$\underline{10
}$&
$\underline{10}$&$22$&$22$&$\underline{9}$&$\underline{1}$&&&&
\\
\hline
$D=11$&$1$&$\underline{11}$&$\underline{41}$&$75$&$76$&$
\underline{52}$&
$\underline{52}$&$76$&$75$&$\underline{41}$&$\underline{11}$&$1
$&&\\
\hline
$D=13$&$1$&$13$&$\underline{64}$&$\underline{168}$&$267$&$2
79$&$\underline{2
32}$&
$\underline{232}$&$279$&$267$&$\underline{168}$&$\underline{64}
$&$13$&1\\
\hline
\end{tabular}
\caption{Number of independent octonionic tensorial charges with 
underlined
 octonionic-Hermitean matrices. The signatures entering the tables 
are respectively given
 by $(0,7), (9,0) or (1,8),(10,1) or (2,9),(11,2) or (3,10)$.}
\end{center}
\end{table}
The Table $1$ is valid for octonionic generalized Poincar\'e
superalgebras, with Abelian generators, as well as for their
nonAbelian counterparts $U_\alpha(k,{\bf O})$ ($k=2$ for $D=9$,
$k=4$ for $D=11$, $k=8$ for $D=13$) describing octonionic
$D$-dimensional AdS superbranes. The octonionic equivalence of
different sectors, via generalized Poincar\'{e} or
generalized AdS supersymmetry algebras  interpreted as
branes sectors,  can be symbolically expressed in different odd
space-time dimensions according to the Table~$2$.
{{\begin{eqnarray}&
\begin{tabular}{|c|c|}\hline
$D=7$& $M0\equiv M3$\\ \hline $D=9$&$M0+M1\equiv M3$\\ \hline
$D=11$&$M1+M2\equiv M5$\\ \hline $D=13$&$M2+M3\equiv 
M6$\\ \hline
\end{tabular}
& \nonumber
\end{eqnarray}}}
\centerline{Table $2$. The relation between octonionic
super-$p$-branes $M_p$.}
\quad\par
 In $D=11$ dimensions the relation between $M1+M2$ and $M5$
can be made explicit as follows. The $11$ vectorial indices $\mu$
are split into the $4$ real indices, labeled by $a,b,c,\ldots$ and
the $7$ octonionic indices labeled by $i,j,k,\ldots$.
 We get:
 {{\begin{eqnarray*}
\begin{tabular}{cc}
$M1 +M2$ & $M5$\\ \\
\begin{tabular}{cc}
$4$& $M1_a$\\
$7$&$M1_i$\\
$6$&$M2_{[ab]}$\\
$4\times 7= 28$&$M2_{[ai]}$\\
$7$& $ M2_{[ij]}\equiv M2_{i}$
\end{tabular}
&
\begin{tabular}{cc}
$7$& $M5_{[abcdi]} \equiv M5_i$\\
$4\times 7=28$&$M5_{[abcij]}\equiv M5_{[ai]}$\\
$6$&$M5_{[abijk]}\equiv M5_{[ab]}$\\
$4$&$M5_{[aijkl]}\equiv {M5}_a$\\
$7$& $ M5_{[ijklm]}\equiv {\widetilde M5}_{i}$
\end{tabular}
\end{tabular}
\end{eqnarray*}}}
which shows the equivalence of the two sectors, as far as the
 octonionic content and
tensorial properties are concerned. Please notice that the correct
total number of $52$ independent components is recovered
\begin{eqnarray*}
52 &=& 2\times 7 +28+6+4.
\end{eqnarray*}

It would be very interesting to find a dynamical realization of
presented above octonionic super-$p$-branes framework. Similarly
one can reproduce the count of independent degrees of freedom for
octonionic $M2, M3 ,M6$ in $D=13$.

\section{Outlook}

The octonions are the basic ingredients of many exceptional
structures in mathematics. It is very well known, that
  the octonions provide the algebraic and geometric framework for 
the
 exceptional Lie algebras.
Indeed, $G_2$ is the automorphism group of the octonions, while
$F_4$ is the automorphism group of the $3\times 3$
octonionic-valued hermitian matrices realizing the exceptional
$J_3({\bf O})$ Jordan algebra. $F_4$ and the remaining exceptional
Lie algebras ($E_6$, $E_7$, $E_8$) are recovered from the
so-called ``magic square Tit's construction" which associates a
Lie algebra to any given pair of division algebras, if one of
these algebras coincide with the octonionic algebra
\cite{bs}.
We would like also to point out here that exceptional Lie
algebras have numerous applications in elementary particle
 physics (see e.g. [21])

We have applied the octonionic structure to the description of a
new version of $M$-theory.
 The main outcome of our considerations, which is
 symbolically represented in Table~2,
implies that in such a framework the different brane sectors are
no longer independent. We would like also to point out (see
formula (2.14)) that
 octonionic structure imposes
 in extended space-times [23,24,19]
 additional constraints on
 central charge tensor coordinates, without restricting
 however  D=11
  spacetime.

   Our considerations here are purely algebraic -
   the step which would be desirable is to provide
   some corresponding geometrical notions.
   It should be pointed out, however, that for nonassociative
   algebras the distinction between algebraic and geometric
   considerations rather disappears.

\subsection*{Acknowledgments}
The authors would like to thank M. Cederwall for discussions.

\end{document}